\documentclass[10pt, conference]{IEEEtran}

\IEEEoverridecommandlockouts
\usepackage{cite}
\usepackage{amsmath,amssymb,amsfonts}
\usepackage{algorithmic}
\usepackage{graphicx}
\usepackage{textcomp}
\usepackage{xcolor}
\usepackage{colortbl}
\usepackage{multicol}
\usepackage{pifont}
\usepackage{multirow}
\usepackage{makecell}
\usepackage{arydshln}
\usepackage{booktabs}
\usepackage{threeparttable}
\usepackage{multirow}
\usepackage{subfigure}
\usepackage{enumitem}

\usepackage{fontawesome}

\def\BibTeX{{\rm B\kern-.05em{\sc i\kern-.025em b}\kern-.08em
    T\kern-.1667em\lower.7ex\hbox{E}\kern-.125emX}}
\begin{document}

\title{Refactoring Deep Learning Code: A Study of Practices and Unsatisfied Tool Needs\\
}

\author{
	\IEEEauthorblockN{ Siqi Wang, Xing Hu*, Bei Wang, Wenxin Yao, Xin Xia, Xinyu Wang\thanks{*Corresponding author.}}
	\IEEEauthorblockA{Zhejiang University, Hangzhou, China} 
    \IEEEauthorblockA{\{siqiwang, xinghu, wenxinyao, wangxinyu\}@zju.edu.cn, bwang9410@gmail.com, xin.xia@acm.org}
}


\maketitle

\begin{abstract}
With the rapid development of deep learning, the implementation of intricate algorithms and substantial data processing has become a standard element of deep learning projects. As a result, the code has become progressively complex as the software evolves, which is difficult to maintain and understand. 
Existing studies have investigated the impact of refactoring on software quality within non-deep learning software. However, the insights of code refactoring in the context of deep learning are still unclear. This study endeavors to fill this knowledge gap by empirically examining the current state of code refactoring in the deep learning realm and practitioners' views on refactoring tools. We first manually analyze the commit history of five popular and well-maintained deep learning projects (e.g., PyTorch). We mine 4,401 refactoring practices in 2,445 historical commits and measure how different types and elements of refactoring operations are distributed.
We then survey 159 practitioners about their views of code refactoring in deep learning projects and their expectations of current refactoring tools. The survey result shows that refactoring research and the development of related tools in the field of deep learning are crucial for improving project maintainability and code quality, and that current refactoring tools do not adequately meet the needs of practitioners. Lastly, we provide our perspective on the future advancement of refactoring tools and offer suggestions for developers' development practices.

\end{abstract}


\section{Introduction}
\label{sec:introduction}
As deep learning continues to evolve rapidly, deep learning projects continue to be rapidly updated to optimize model construction, improve computing, and increase algorithm performance~\cite{guo2019empirical,whang2020data}. However, if maintenance activities are not conducted properly, they can lead to a decrease in quality. The complexity of deep learning models and their high dependence on data, as well as constantly updated algorithms and techniques, present unique challenges for their maintenance. These unique challenges make the development and maintenance of deep learning projects distinct from traditional software~\cite{amershi2019software}.

There have been significant studies ~\cite{fowler2018refactoring, harman2007pareto,lin2016interactive,meananeatra2012identifying,alomar2022refactoring, alomar2019impact} demonstrating the benefits of refactoring for software maintenance, reuse, and code enhancement, focused on traditional software. 
However, little attention has been given to refactoring in DL projects or practitioners’ perspectives on refactoring tools. Existing approaches largely target traditional software, overlooking DL-specific characteristics such as experiment-driven workflows and configuration-heavy designs.
As a result, current tools may misinterpret common deep learning practices as code smells, leading to poor adoption and increased maintenance costs.
Investigating code refactoring in deep learning projects and uncovering the reasons behind such practices can help optimize the development process, improve team productivity, and enhance code quality. Therefore, we analyze the state of code refactoring in deep learning repositories and investigate the perceptions of deep learning practitioners on code refactoring.

Our study aims to answer the following research questions:

\textbf{RQ1:} \textbf{How does code refactoring behave within deep learning projects?} 
    
This RQ studies code refactoring practices in deep learning projects and uncovers the distribution of different refactoring operation types and elements’ usage in their projects.
This RQ highlights the unique challenges and implications within the deep learning domain. It also lay the groundwork for tailored software engineering that is crucial to the rapidly evolving field of deep learning. s
Additionally, the manual examination of commit messages unveils indications of developers employing automation tools for refactoring tasks. Based on this finding, we further conduct a survey to investigate the compatibility of existing tools with the unique needs of deep learning practitioners in RQ3.

\textbf{RQ2:} \textbf{What are the perspectives of deep learning practitioners regarding code refactoring?} 
    
Building upon insights gained from RQ1, this RQ investigates practitioners' perspectives on code refactoring, including their opinions on specific refactoring operations and elements within deep learning. 
Examining practitioners' perspectives on deep learning code refactoring can validate and quantify the observations from RQ1, and provide a critical understanding of their preferences, challenges, and potential unmet needs. These insights help bridge the gap between the research and practice in current AI development, thereby aiding the development of targeted refactoring techniques for this domain.
    
\textbf{RQ3:} \textbf{How well do current refactoring tools meet practitioner needs?} 
   
   This RQ investigates practitioners' opinions on the effectiveness of existing refactoring tools.
   The findings from RQ1 reveal evidence of refactoring tool utilization within commit messages, which leads us to assess the suitability of existing tools for deep learning requirements.
   This RQ aims to inform the development of future tools tailored to the distinct requirements of deep learning practitioners.

The intention behind our investigation is to facilitate consideration by researchers regarding the requirements of practitioners, thereby continuing the advancement of code refactoring for deep learning projects. Furthermore, we aim to provide new insights that can promote the development of better refactoring tools for deep learning projects. This paper makes the following contributions:

\begin{enumerate}[leftmargin=*]
    \item We manually analyzed five deep learning projects' commits and identified 27 types of 4,401 refactoring practices. We further analyzed the distribution of different refactoring operation types and elements in deep learning projects.
    \item We surveyed 159 deep learning practitioners from 38 countries to shed light on practitioners' views on refactoring and their expectations of refactoring tools. To the best of our knowledge, we are the first to perform an empirical study of refactoring practices in deep learning projects. 
\end{enumerate}

\noindent\textbf{Paper Organization:} Section \ref{sec:methodology} describes the methodology of our research. Section \ref{sec:results} shows the results of our study. We discuss the implications and threats of our results in Section \ref{sec:discussion}. Section \ref{sec:relatedwork} discusses related work. Section \ref{sec:conclusion} draws conclusions and outlines avenues for future work.

\section{Research Methodology}
\label{sec:methodology}

\begin{figure}
	\centering
	\includegraphics[width=.8\linewidth]{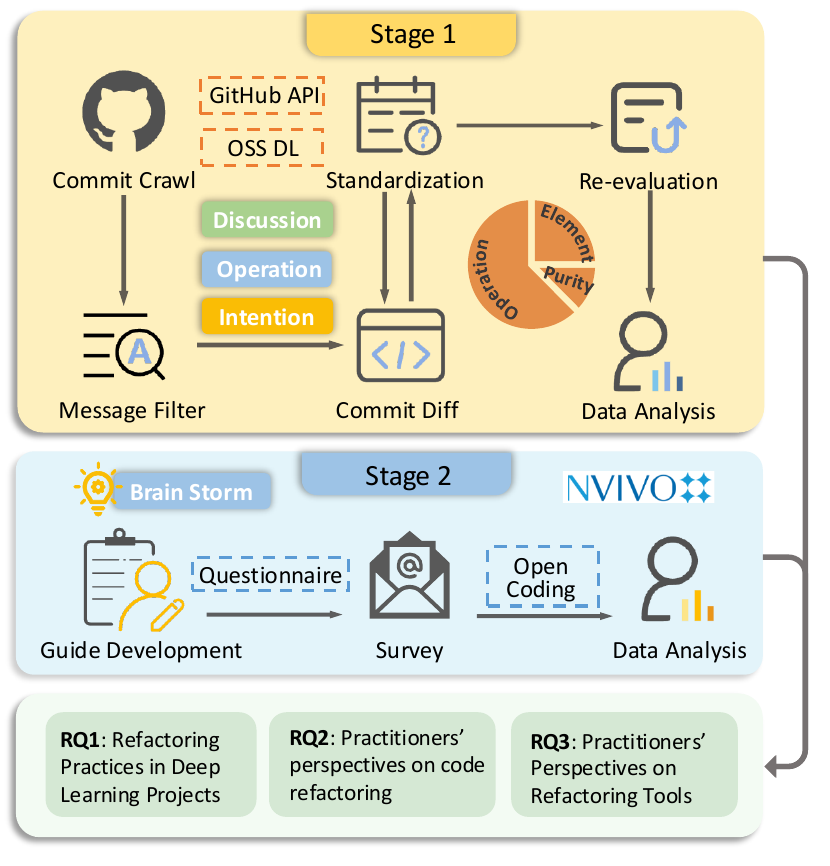}
	\caption{Research Methodology Overview}
	\label{figure-overview}
\end{figure}

In this section, we present the design of our empirical study. Our main goal in this study is to comprehensively understand code refactoring practices within deep learning projects, investigate practitioners' perceptions of code refactoring, and assess the alignment of current refactoring tools with the specific needs of deep learning practitioners. The overview of the methodology in our study is shown in Figure~\ref{figure-overview} and consists of two stages. \textbf{Stage 1:} Manual mining of refactoring operations from repository history commit messages manually. \textbf{Stage 2:} An online survey for confirming and extending the conclusions about the current stage of refactoring in open source deep learning libraries.

\subsection{Stage 1: Refactoring Manual Detection}

Since Python is the most popular programming language for deep learning due to its readability, extensive libraries, and frameworks~\cite{raschka2017python,ketkar2017deep,saitoh2021deep}, we only analyze refactoring commits that involve Python. To answer RQ1, we mine the commits of five open source deep learning frameworks that are widely used and well-maintained: Keras~\cite{Keras}, Scikit-Learn~\cite{Scikit-learn}, PyTorch~\cite{Pytorch}, TensorFlow~\cite{Tensorflow}, and Transformer~\cite{Transformers}. These projects are chosen based on their high stars in GitHub.
Table \ref{tab-statistic} shows statistics of the projects we use in RQ1. To the best of our knowledge, no refactoring detection tool can detect all common refactoring operations in Python. Additionally, some commits might contain tangled changes, making it more difficult to isolate the changes related to refactorings. Therefore, we follow previous research to manually mine refactoring operations from the commit~\cite{alomar2019can, kim2014empirical, pantiuchina2020developers ,silva2017refdiff}. Additionally, we use PyRef~\cite{atwi2021pyref} to assist in the detection of specific operations, such as \textit{Rename Method}, \textit{Rename Parameter}, and \textit{Add Parameter}, with the final results still determined manually.

\begin{table}[]
\centering
\caption{Statistics of deep learning projects used in our study.}
\label{tab-statistic}
\resizebox{0.88\linewidth}{!}{%
\renewcommand{\arraystretch}{1.2}
\begin{tabular}{llcc}
\toprule
\textbf{Project}      &\textbf{Domain}       & \textbf{\#Commits}  &\textbf{\#Star} \\ \midrule
Keras        & Data Science, JAX & 8,342 & 60.2k          \\ 
Scikit-learn & Data Analysis & 30,375  & 57k       \\ 
PyTorch      & GPU, Neural-network & 63,722  & 74.3k       \\ 
Transformers & NLP, Pretrain-model & 13,900  & 118k       \\ 
Tensorflow   & Tensor, Machine-learning & 50,863  & 180k       \\ \midrule
\textbf{Total} &               & 167,202 & \\

\bottomrule
\end{tabular}
}
\end{table}
\vspace{-1pt}

\subsubsection{Filter}
We first crawl all 167,202 commits of these five deep learning projects using the GitHub API. However, due to the enormous amount of manual detection, we use a keyword-based filter to reduce the amount of work involved in manual detection. 
In this part, following \textit{Self-Affirmed Refactoring}~\cite{alomar2019can}, we pre-filter potential refactoring commits with a keyword-based commit message filter, including \textit{`refactor'}, \textit{`clean up'}, and \textit{`reorganize'}, etc. In particular, we only focus on refactoring commits in Python since the majority of the deep learning project is written in Python~\cite{raschka2017python}. This process reduced the candidate set to 28,803 commits.

\subsubsection{Manual Detection}

Next, we conducted a two-stage manual identification process to determine whether each candidate commit indeed involved a structural refactoring: 

\ding{192}The 28,803 filtered commits were evenly divided between two authors, who independently examined each commit message. Commits clearly unrelated to refactoring (e.g., messages starting with \textit{[fix]}, \textit{[test]}) were immediately labeled as non-refactoring and excluded.
\ding{193}For the remaining commits, the evaluators further analyze the code changes of the refactoring commit and the parent commit to determine what refactoring operation is performed by that commit, we follow Fowler's description~\cite{fowler2018refactoring}. The evaluators ignore the tangled changes in the manual detection phase and only abstract refactoring related changes. Since refactoring practices in real-world development may differ from the description given in Fowler's book, after each evaluator has marked 100 commits, they have a discussion to re-establish consistent refactoring classification standards. The details of the standard are available in the replication package~\cite{Replication_Package}. The evaluators repeat this action in all five studied projects and correct each of the previously marked commits according to the final standards. For each commit, we extract information modeling, including the following elements:

\begin{itemize}[leftmargin=*]
    \item \textbf{Module:} A module is defined as a Python file containing classes and methods and directly affiliated definitions and statements (i.e., those not defined within a method).
    \item \textbf{Class:} A class may contain attributes, methods, and statements.
    \item \textbf{Method:} A method contains a list of parameters, and the statements it carries.
    \item \textbf{Statements:} A statement is the smallest execution unit in Python and is terminated by a line break or a semicolon.
    \item \textbf{Variable:} A variable is a container for storing data and is the smallest nameable storage element in python program.
\end{itemize}

The eight groups of refactoring operation types include the six most common refactoring operation types in deep learning projects and two refactoring operation types (\textit{Pull Up} and \textit{Push Down}) that are widely studied in previous work~\cite{atwi2021pyref, brito2021characterizing}.

\begin{itemize}[leftmargin=*]
    \item \textbf{Clean Up Refactoring:} Removing invalid or redundant code fragments that are no longer being executed by the program. This type of refactoring is sometimes accompanied by words like \textit{``clean up duplicated''} and \textit{``remove unused code''} in the commit message.
    \item \textbf{Rename Refactoring:} Changing the name of elements for code clarity and readability. This type of refactoring consists of \textit{Rename Module}, \textit{Rename Class}, \textit{Rename Method}, \textit{Rename Variable}, and \textit{Rename Parameter}.
    \item \textbf{Move Refactoring:} Moving code elements from one location to another. This type of refactoring consists of \textit{Move Module}, \textit{Move Class}, \textit{Move Method}, \textit{Move Statement}, and \textit{Move Variable}.
    \item \textbf{Extract Refactoring:} Extracting independent and reusable elements from more complex code elements. This type of refactoring consists of \textit{Extract Module}, \textit{Extract Class}, \textit{Extract Method}, and \textit{Extract Variable}. 
    \item \textbf{Inline Refactoring:} Moving a part of code directly into its place of use to replace the call for these codes. This type of refactoring consists of \textit{Inline Module}, \textit{Inline Class}, \textit{Inline Method}, and \textit{Inline Variable}.
    \item \textbf{API Refactoring:} Refactoring the application programming interfaces in code. This type of refactoring usually involves parameter additions, deletions and changes, etc.
    \item \textbf{Push Down Refactoring:} Redistributing/changing code elements from superclass to subclass. This type of refactoring consists of \textit{Push Down Method} and \textit{Push Down Class}. 
    \item \textbf{Pull Up Refactoring:} Redistributing/changing code elements from subclass to superclass. This type of refactoring consists of \textit{Pull Up Method}.
\end{itemize}

Only commits with both an explicit refactoring intent and concrete structural code changes were retained. After the independent evaluation, the two evaluators check each other's data that is labeled as refactoring-related. The evaluators disagree on 181 instances out of a total of 4,401 refactoring instances. These instances are carefully discussed to reach a consensus until there is no disagreement on any code change after the second stage of refactoring operation classification. The inter-rater agreement is further quantified using Cohen's Kappa coefficient, yielding a value of 0.95, indicating substantial agreement between the evaluators in the classification of refactoring operations. 

\begin{figure}
    \centering
    \includegraphics[width=0.9\linewidth]{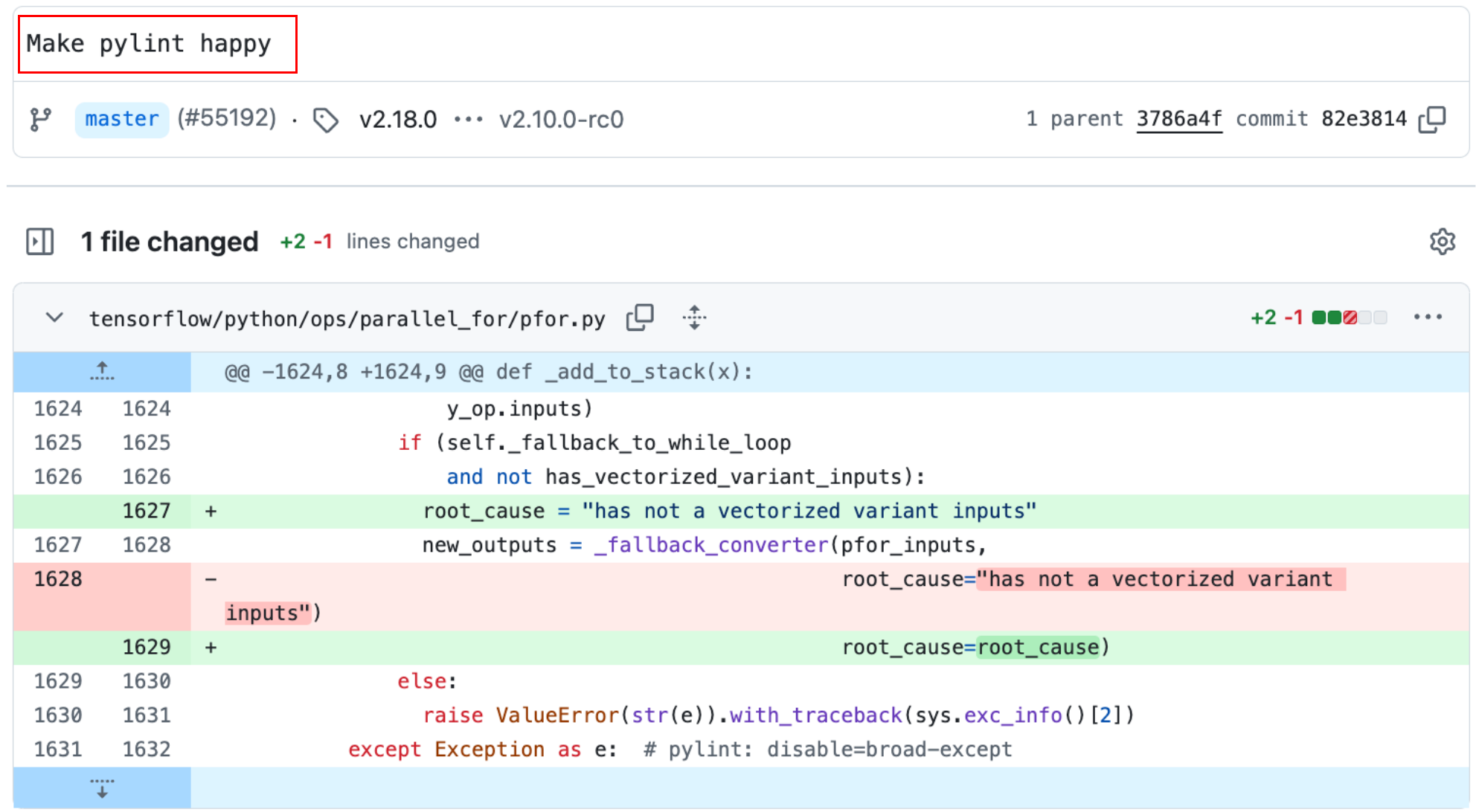}
    \caption{Refactoring Tool Usage in Commit Message}
    \label{fig:tool_message}
\end{figure}
 The labeled dataset is available in replication package~\cite{Replication_Package}. Additionally, we have observed instances of developers mentioning refactoring tools (e.g., ``Pylint'', ``Flake'', and ``Generated by Copilot'') in commit messages. As shown in figure~\ref{fig:tool_message}, the string ``\texttt{has not a vectorized variant inputs}'' in commit 82e3814bc57fb29236ca703329374455f65f672f is extracted as a new variable `\texttt{root\_cause}' to `\texttt{Make Pylint Happy}'. This observation highlights developers' reliance on automated refactoring tools, motivating us to further explore practitioners' perspectives on these tools in \textbf{Stage 2}.

\subsection{Stage 2: Online Survey}

To further investigate deep learning code refactoring, we conduct an anonymous online survey with deep learning participants. The survey aims to validate and quantify the observations from Stage 1, and to shed light on practitioners' views on refactoring and their expectations of refactoring tools.

\subsubsection{Design}
Combining the results of the first step and previous work~\cite{atwi2021pyref,brito2021characterizing}, we conclude eight refactoring operation types, five refactoring operation elements, and two types of refactoring tools to launch an online survey for deep learning practitioners. The online survey aims to provide insights into open source code refactoring in deep learning projects and practitioners' expectations of current refactoring tools. 
The refactoring operation types we employed come from the eight refactoring operation types from Stage 1 consist of \textit{Clean Up, Rename, Move, Extract, Inline, API, Pull Up, and Push Down} refactoring. The elements consist of \textit{Variable, Statement, Method, Class}, and \textit{Module}. The refactoring tools consist of code smell detection tools and automatic refactoring tools~\cite{Survey}.

The survey include different types of questions, e.g., multiple-choice questions, short answer questions, and rating questions (5-point Likert scale: Strongly Disagree to Strongly Agree). We include the category ``I don't know'' to filter respondents who do not understand our brief descriptions. The survey consists of three sections:
\begin{itemize}[leftmargin=*]
    \item \textbf{Demographics:} The survey first asks for demographic information about the participants, including country/area of residence, current occupation, experience in years, and primary programming language.
    \item \textbf{Thoughts on Refactoring:} This section focuses on providing insights into the current state of open source code refactoring in deep learning projects. We start by showing clear and concise example diagrams of some of the classic refactoring operations in method level. We then invite developers to rate the eight refactoring operations and five refactoring elements in terms of importance and frequency. This section highlights practitioners' opinions towards refactoring in the deep learning software development process.
    \item \textbf{Tool Utilization:} The purpose of this section is to gather information about developers' usage of refactoring tools and investigate practitioners' expectations of these tools. We first provide respondents with a brief description of code refactoring tools shown in Table \ref{tab-tool description}, consisting of (1) code smell detection tool~\cite{guo2010domain} and (2) automatic refactoring tools~\cite{griffith2011truerefactor}. Then we ask practitioners \textit{Have you used or are you familiar with such tools?} and extend an invitation to respondents to share their observations regarding such tools. In addition, practitioners are invited to provide advice on improving refactoring tools through an open-ended question.
\end{itemize}


\begin{table}
\caption{Description of Code Refactoring Tools}

	\label{tab-tool description}
\renewcommand{\arraystretch}{1.2}
\centering
\resizebox{0.95\linewidth}{!}{%

\begin{tabular}{cl}
\toprule
\textbf{Refactoring Tool} &
  \multicolumn{1}{c}{\textbf{Description}} \\ \toprule

\parbox[c]{2cm}{Code Smell Detection Tools} &
\parbox[c]{6cm}{\vspace{1mm}Code smells are signs that your code is not as clean and maintainable as it could be. They can derive from the misuse of syntax and almost always suggest code needs to be refactored or redesigned to improve the overall quality of the program. Code smell detection tools can help developers find where to refactor to improve the quality of their code.}  \vspace{1mm}\\ \midrule 

\parbox[c]{2cm}{Automatic Refactoring Tools} &
  \parbox[c]{6cm}{\vspace{1mm}Automatic refactoring tools identify problems in the code and eliminate them through refactoring. These tools reduce the effort of developers as they have very little to do during the code refactoring process.}\vspace{1mm} \\ \bottomrule
\end{tabular}
}

\end{table}

At the end of the survey, we allow respondents to provide free-text comments, suggestions, and opinions about code refactoring and our survey. A respondent may or may not provide any final comments.

 During the initial phase of our research, we conduct a preliminary survey with a small group of professionals who differed from our survey respondents. The purpose is to gather feedback on two key aspects: (1) the length of the questionnaire and (2) the clarity of the terms used. Based on the feedback received, we make minor modifications to the survey and finalize the questionnaire. It should be noted that the responses collected during the pilot survey are not included in the results presented in our research paper.

 \begin{table}[h]
\centering
 \caption{Participants Roles \& Programming Experience}
\label{tab-demographics}
\resizebox{0.88\linewidth}{!}{
\small
\renewcommand{\arraystretch}{1.2}
\begin{tabular}{ccccccc}
\hline
                                                 & 0-1 y & 2-3 y & 4-5 y & 6-9 y & \textgreater{}10 y & total \\ \hline
{\color[HTML]{1F1F1F} Algorithm} & 4     & 25    & 26    & 14    & 7                  & 76    \\
Development                                      & 5     & 14    & 10    & 15    & 7                  & 51    \\
Architect                                        & 0     & 1     & 2     & 3     & 5                  & 11    \\
Project   Manager                                & 1     & 1     & 0     & 1     & 1                  & 4     \\
Testing                                          & 0     & 1     & 0     & 1     & 1                  & 3     \\
Others                                           & 2     & 7     & 2     & 2     & 1                  & 14    \\ \hline
total                                            & 12    & 49    & 40    & 36    & 22                 & 159   \\ \hline
\end{tabular}
}
\end{table}

\subsubsection{Participant Recruitment}
We select Github repositories with the top 100 popular open-source deep learning projects (based on their number of stars) and mine these repositories to extract their contributors' public email addresses. We finally mine 3,125 contributors' email addresses and sent a link to our survey. We aim to recruit open-source deep learning practitioners who have software development experience in addition to professionals working in the industry. Out of these emails, four practitioners replied with blank responses; six practitioners replied that they would not answer any survey. Finally, we received 159 valid responses. The 159 respondents resided in 38 countries across six continents. The top two countries where the respondents came from are India and the United States.

An overview of the surveyed participants and their experience is depicted in Table ~\ref{tab-demographics}. Most participants are engaged in Algorithm and have 2-3 years of professional experience.

\subsubsection{Data Analysis}
We analyze the survey results based on the question types.

To understand trends in the Likert-scale questions, we report the percentage of each option selected. We drop ``I don't know'' ratings and create bar charts (many of which are shown in the remainder of this paper). 

To obtain insights from responses to open-ended questions, we use open coding to analyze the survey results qualitatively by inspecting responses. The first author analyzes the survey respondents by transcribing them and then performs open coding to generate codes of the questionnaire contents using NVivo~\cite{NVivo} qualitative analysis software. Then, the second author verifies the initial codes created by the first author and provides suggestions for improvement.

\vspace{-1pt}
\section{Results}
\label{sec:results}

This section presents the results of research into code refactoring from commits and practitioners.


\subsection{RQ1: Refactoring Practices in Deep Learning Projects}
\label{subsec:3.1}

In RQ1, we explore code refactoring practices in deep learning projects, including practitioners' practices on refactoring during development, and the distribution of different refactoring operation types and elements' usage in their projects.

\begin{table}[]
\caption{Manual Detection Results}
    \centering
\renewcommand{\arraystretch}{1.2}
\resizebox{.95\linewidth}{!}{
\begin{tabular}{clcccc}
\toprule
\rowcolor[gray]{0.85}
\multicolumn{1}{c}{\textbf{Group}} & \multicolumn{1}{c}{\textbf{Refactoring operation}} & \multicolumn{1}{c}{\textbf{\#Pure}} & \multicolumn{1}{c}{\textbf{\#Multi}} & \multicolumn{1}{c}{\textbf{\#Non-ref}} & \multicolumn{1}{c}{\textbf{\#Total}} 
\\ \toprule
 Clean Up                    & Clean Up Refactoring                  & 990                       & 71                           & 155                          & 1,216                      
\\ \midrule

\rowcolor[gray]{0.93}
    & Rename Variable                            & 214                       & 44              & 86                           & 344                        \\
\rowcolor[gray]{0.93}             & Rename Module                              & 119                       & 20                           & 22                           & 161                        \\
 \rowcolor[gray]{0.93}  & Rename Method                              & 320                       & 129                          & 136                          & 585                        \\
\rowcolor[gray]{0.93} & Rename Class                               & 93                        & 49                           & 38                           & 180                        
  \\ 
  \rowcolor[gray]{0.93} \multirow{-5}*{\shortstack{Rename\\ \scriptsize (1,451)}}   
    & Rename Parameter                           & 88                      & 43  & 50         & 181  \\ \midrule
    
   \multirow{5}{*}{\shortstack{API\\\scriptsize (367)}}                           & Reorder Parameter                          & 6                         & 2                            & 0                            & 8                          \\
                            & Remove Parameter                           & 137                       & 51                           & 33                           & 221                        \\
                            & Change Parameter                           & 8                         & 4                            & 4                            & 16                         \\
                            & Add Parameter                              & 48                        & 42                           & 26                           & 116                        \\
                            & Merge Parameter                            & 1                         & 4                            & 1                            & 6                          \\ \midrule
\rowcolor[gray]{0.93}       & Move Variable                              & 35                        & 11                           & 5                            & 51                         \\
 \rowcolor[gray]{0.93}                           & Move Statement                             & 53                        & 12                           & 13                           & 78                         \\
 \rowcolor[gray]{0.93}                           & Move Module                                & 155                       & 21                           & 11                           & 187                        \\
 \rowcolor[gray]{0.93}                           & Move Method                                & 158                       & 74                           & 36                           & 268                        \\
 \rowcolor[gray]{0.93}   \multirow{-5}*{\shortstack{Move\\\scriptsize (630)}}                        & Move Class                                 & 22                        & 19                           & 5                            & 46                         \\ \midrule
     & Push Down Method                           & 1                         & 0                            & 0                            & 1                          \\
                         & Push Down Class                            & 2                         & 1                            & 0                            & 3                          \\
      \multirow{-3}*{\shortstack{Push\&Pull\\ \scriptsize (10)}}                      & Pull Up Method                             & 2                         & 2                            & 2                            & 6                           \\ \midrule
  \rowcolor[gray]{0.93}    & Inline Variable                            & 24                        & 3                            & 6                            & 33                         \\
   \rowcolor[gray]{0.93}                          & Inline Module                              & 15                        & 2                            & 2                            & 19                         \\
    \rowcolor[gray]{0.93}                         & Inline Method                              & 30                        & 6                            & 4                            & 40                         \\
  \rowcolor[gray]{0.93} \multirow{-4}*{\shortstack{Inline\\ \scriptsize (97)}}                           & Inline Class                               & 4                         & 1                            & 0                            & 5                          \\ \midrule
\multirow{4}{*}{\shortstack{Extract\\ \scriptsize (630)}}    & Extract Variable                           & 26                        & 6                            & 21                           & 53                         \\
                            & Extract Module                             & 212                       & 31                           & 16                           & 259                        \\
                            & Extract Method                             & 135                       & 44                           & 72                           & 251                        \\
                            & Extract Class                              & 45                        & 12                           & 10                           & 67                         \\ \midrule

 \rowcolor[gray]{0.85}\textbf{Total}  &                                            & 2,943                      & 704                          & 754                          & 4,401       \\   \bottomrule 

\end{tabular}}

\begin{quote}
    \footnotesize{*\textit{Pure}: only one refactoring type practices without tangle change; \textit{Multi}: practices with more than one refactoring operation type without fix or update; \textit{Non-ref}: refactoring practices with other changes including fix and update. }
\end{quote}

\label{tab-manul detection}
\end{table}

\subsubsection{Refactoring Operations}

We identify 27 refactoring operations of 4,401 practices by manual detection, the complete dataset can be found in the replication package~\cite{Replication_Package}. To enhance comprehension and organization of these refactorings, we categorize common refactoring operations into eight group operation types. \textit{Pull Up} and \textit{Push Down} refactoring are combined and displayed because they both involve inheritance relationships of code elements and the number is small.
Table \ref{tab-manul detection} shows the result of our manual detection, including the refactoring practice number, and whether the practices are pure or contain tangled changes. 

According to the table, \textit{Clean Up Refactoring} is one of the most frequent refactoring operation types with 1,216 practices. This high frequency reflects the iterative nature of deep learning projects, where rapid prototyping often leads to the accumulation of outdated or redundant code, as developers focus on testing and adjusting model parameters or algorithmic choices. For example, many commit messages include phrases like ``\textit{remove the old code...}'' as out-time code will be discarded once it is updated swiftly. Besides, deep learning projects usually include experiments, such as testing new models, and algorithms or tuning hyper-parameters~\cite{amershi2019software} as ``\textit{remove test of model...}''. This leads to a large number of code snippets in the source, that ultimately become dead code when the outputs of the experiment are determined and subsequently thrown away. Nevertheless, as requirements can change over time or the originally intended functionality not be implemented, a significant amount of dead code is left behind. \textit{Clean Up Refactoring} is an important step in maintaining a clear code structure and improving the maintainability of projects~\cite{romano2018multi}.

\textit{Rename} operations (including renaming variables, modules, methods,  classes, and parameters) are also a common type of refactoring, with a total of 1,451 practices. In particular, \textit{Rename Method} and \textit{Rename Variable} had the highest number of practices, with 585 and 344 respectively. This may be due to the fact that deep learning projects usually contain a large amount of iterative code, and developers are constantly optimizing the code for readability and maintainability. Renaming methods and variables, in particular, helps to make the code more understandable and reduces the coupling between codes.

 \textit{Move} is also commonly used in the refactoring of deep learning projects with 630 practices. Notably, more practices are observed for \textit{Move Method} and \textit{Move Module}, with 268 and 187 commits. This indicates that developers frequently relocate code units during refactoring in order to enhance the code structure's organization. \textit{API Refactoring} and \textit{Extract} have 351 and 356 practices. Model structures and algorithms in deep learning projects can often be complicated, requiring more abstraction and optimization. Consequently, \textit{Extract} is frequently employed to abstract complex methods or modules, thereby enhancing the modularity and reusability of code. Deep learning projects always need to refine and enhance their interfaces regularly to accommodate evolving needs, resulting in frequent usage of \textit{API Refactoring}~\cite{pantiuchina2020developers}.
 
\textit{Inline Method} is employed in 40 practices, while \textit{Inline Variable} is used in 33 practices. Despite being used less frequently in deep learning projects, these refactoring operations are of particular importance in specific scenarios. \textit{Inline} is typically used to optimize performance and reduce the overheads associated with function calls. \textit{Pull Up} and \textit{Push Down} operations are infrequent in deep learning projects. In the context of deep learning projects, there is a tendency to prioritize the design of model hierarchies and structures, with less emphasis placed on inheritance and optimizing hierarchies.

\begin{center}
    \resizebox{\linewidth}{!}{
\begin{tabular}{l!{\vrule width 1pt}p{0.92\columnwidth}}
    \makecell{{\LARGE \faLightbulbO}}  &\textbf{Finding 1.} 
    \textit{Clean Up Refactoring and Rename are two of the most frequent operation types in deep learning repositories. The high frequency of API Refactoring reflects the importance of interface design and modification in deep learning projects.}\\
\end{tabular}}
\end{center}

\subsubsection{Refactoring elements}
We count the elements of all refactoring practices including \textit{Variable}, \textit{Statement}, \textit{Method}, \textit{Class}, and \textit{Module}.  This distribution of refactoring practice elements can offer insights into code optimization and project maintenance in deep learning projects.


\textit{Method} (44.21\%) is the most common element, include techniques like \textit{Rename Method} and \textit{Extract Method}. These types of refactoring operations are relatively simple and contribute to both code readability and modularity. It is essential to consider these factors when undertaking code maintenance tasks. \textit{Variable} refactoring (19.01\%) has the second-highest percentage of refactoring operations. The most frequent of these is \textit{Rename Variable}, which is typically linked to the ``\textit{same variable name}'' in the commit message to prevent ``\textit{naming conflicts}''. \textit{Module} refactoring (14.22\%) is moderate, with the majority of instances occurring within the context of large-scale refactoring, renaming, and moving module code files. The proportions of \textit{Class} refactoring (10.56\%) and \textit{Statement} refactoring (11.99\%) are similar. Code refactoring in deep learning projects is not limited to one level but needs to be considered at several levels. \textit{Class} refactoring helps improve the overall structure and maintainability of the code, while \textit{Statement} refactoring directly affects the execution efficiency and performance of the code. The two complement each other and together improve the code quality of the project.

The refactoring element chosen by developers reveals a careful balance between risks and benefits. The frequent focus on \textit{Method} and \textit{Variable} refactoring indicates that developers prefer smaller, targeted code changes. This approach offers several advantages: it minimizes risk, limits the scope of impact, and allows for incremental improvements in code quality. In contrast, \textit{Class} and \textit{Module} refactorings, while offering significant potential to improve code organization, involve larger modifications. As a result, developers may approach these more cautiously due to the greater structural changes and potential impact on the codebase.

\begin{center}
    \resizebox{\linewidth}{!}{
\begin{tabular}{l!{\vrule width 1pt}p{0.98\columnwidth}}
    \makecell{{\LARGE \faLightbulbO}}  &\textbf{Finding 2.} 
    \textit{ High frequencies of Variable and Method refactoring suggest that developers focus on data processing and implementation. The lower Class refactoring frequencies may be due to the simple code structure of deep learning. }\\
\end{tabular}}
\end{center}

\subsubsection{Pure refactoring analysis}
In real-world software development, developers may do multiple refactorings in a single commit or mix refactorings with non-refactorings such as `debug' and `update'. We count the purity of refactoring practices, including (1) \textit{Pure} refactoring, (2) \textit{Multi}-type refactoring, which means different types of refactoring operations contained in a single commit, and (3) \textit{Non-ref} refactoring, which is refactoring practice with other non-refactoring operations.

A significant majority of the practices (66.88\%) are purely refactoring operations. This high proportion of pure refactoring indicates that developers are focused on improving code quality without mixing other types of changes. This behavior facilitates the maintenance of clear commit histories.

Additionally, there are also some refactoring practices (15.98\%) that occur in combination with other types of refactoring operations. The most frequent combination of refactoring operations is the \textit{Rename Class} and \textit{Rename Method}, which occurs 52 times. Another frequent combination is \textit{Clean Up Refactoring} and \textit{Rename Method}, which occurs 31 times. The presence of multiple refactoring types within a single commit indicates the implementation of a comprehensive refactoring session, intending to simultaneously enhance various aspects of the code. This approach may prove an efficient means of addressing multiple code quality issues in a single effort, although it may complicate the commit history.

Some practices (17.14\%) are combined with other operations, such as debugging or updating, suggesting a more integrated approach to code maintenance and development. This is a practical approach, but it may also make it more challenging to identify and comprehend each change's impact.

\begin{center}
    \resizebox{\linewidth}{!}{
\begin{tabular}{l!{\vrule width 1pt}p{0.98\columnwidth}}
    \makecell{{\LARGE \faLightbulbO}}  &\textbf{Finding 3.} 
    \textit{ A lot of Pure Refactoring practices evince a commitment to enhancing the quality of the code without incorporating other types of modifications. This practice facilitates more effective change tracking and comprehension of the impact of modifications on the code base.}\\
\end{tabular}}
\end{center}

\subsection{RQ2: Practitioners' Opinion on Refactoring}
In RQ2, deep learning practitioners are surveyed and asked to evaluate the significance of operations or elements during the refactoring process. Of the 159 practitioners surveyed, 145 (91.2\%) ``Strongly Agreed'' or ``Agreed'' that \textit{``refactoring is an important part of software development in deep learning projects''}, 11 chose ``Neutral'', while the remaining three practitioners ``Strongly Disagree'' with the importance of refactoring.



\subsubsection{Refactoring Operations}
\begin{figure}
    \centering
    \includegraphics[width=0.85\linewidth]{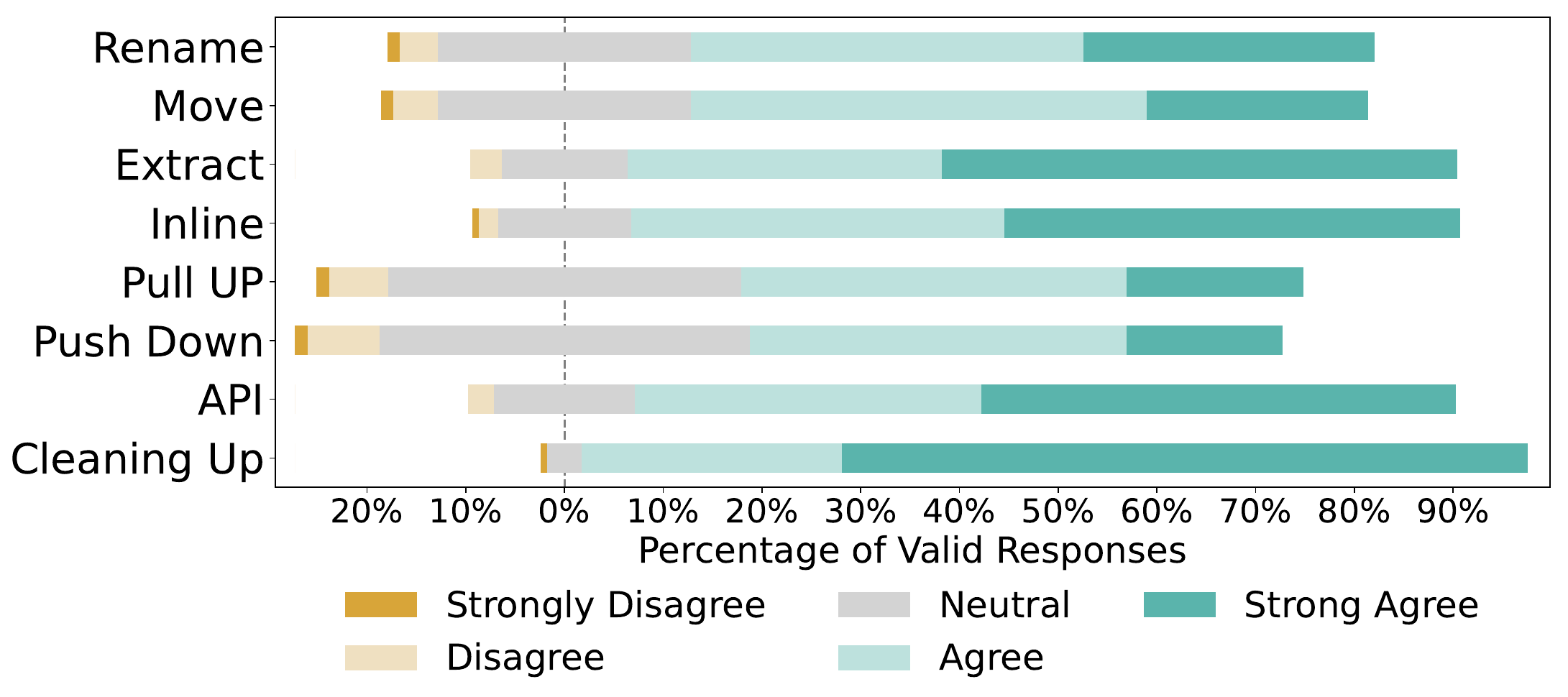}
    \caption{Respondents' Rate of Refactoring Operation Types' Importance}
    \label{figure-operation impact}
\end{figure}

\begin{figure}
    \centering
    \includegraphics[width=0.85\linewidth]{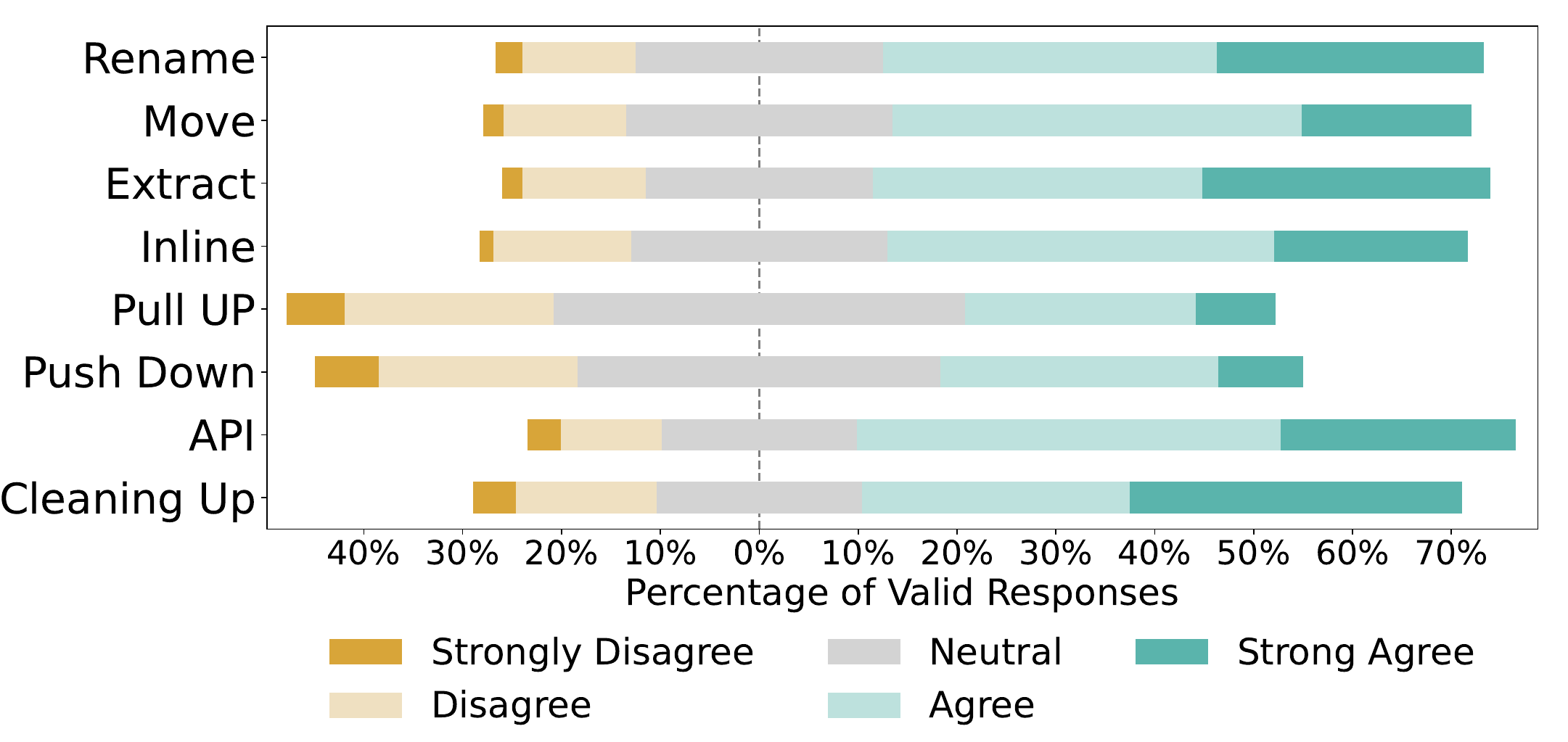}
    \caption{Respondents' Rate of Refactoring Operation Types' Frequency}
    \label{figure-operation freq}
\end{figure}

We further investigate practitioners' opinions on several common refactoring operation types. Figure \ref{figure-operation impact} and Figure \ref{figure-operation freq} illustrate respondents' rating of refactoring operation types' importance and frequency.

In general,  \textit{Clean Up} refactoring emerges as the most vital operation recognized by practitioners. \textit{API Refactoring}, \textit{Extract}, and \textit{Inline} refactorings show similar levels of importance, while \textit{Rename} does not offer significant advantages over \textit{Move}. Additionally, \textit{Pull Up} and \textit{Push Down} are comparable in significance among practitioners. It is worth noting that the practitioners' scores for refactoring operations' frequency are not consistent with our observation in RQ1. There are various possible explanations for this, the one that comes from our dataset used in RQ1 will be discussed in \ref{subsec:Threats}.

The characterization of the rating results for each refactoring operation provides additional insights. As highlighted in RQ1, \textit{Clean Up} is the most frequently applied operation in refactoring commits. However, its perceived importance among practitioners does not reflect this frequency. This is likely because \textit{Clean Up} is often viewed as part of other code changes rather than being a deliberate, stand-alone action. As a result, practitioners may not consciously recognize or prioritize it.
 On the other hand, \textit{API} Refactoring is considered important due to the significance of code interfaces and organizational structure in deep learning projects that typically entail extensive data processing and model creation. However, in RQ1, we do not observe the expected higher frequency of API refactoring, possibly due to the fact that we exclude many API changes involving functionality changes during the manual detection process.
 \textit{Extract} and \textit{Inline} refactoring operations that are important in other languages are also of high importance in the minds of deep learning practitioners, reflecting their concern for code reusability and structural clarity. while \textit{Move} and \textit{Rename} receive relatively low scores, perhaps because they are also relatively low in difficulty to perform, and so receive practitioner slights. In contrast to the above refactoring operations, practitioners tend not to prioritize \textit{Pull Up} and \textit{Push Down}. Developers rank these two actions lowest in terms of importance and frequency of usage, consistent with our observations and discussions in RQ1. We also find an interesting phenomenon when analyzing the data, that more experienced practitioners (who have been working for a longer period of time) are less likely to agree with the importance of these two refactoring operations for code refactoring in deep learning projects.

\begin{center}
    \resizebox{\linewidth}{!}{
\begin{tabular}{l!{\vrule width 1pt}p{0.98\columnwidth}}
    \makecell{{\LARGE \faLightbulbO}}  &
\textbf{Finding 4}:
\textit{According to practitioners' perceptions, \textit{Clean Up}  Refactoring is considered highly important and \textit{API} is most frequently employed refactoring operation, while operations like \textit{Rename}, \textit{Extract}, \textit{Inline}, and \textit{Move} are less valued in deep learning projects. Additionally, \textit{Pull Up} and \textit{Push Down} receive particularly low recognition.}
\end{tabular}}
\end{center}

\subsubsection{Refactoring elements}
\begin{figure}
    \centering
    \includegraphics[width=.75\linewidth]{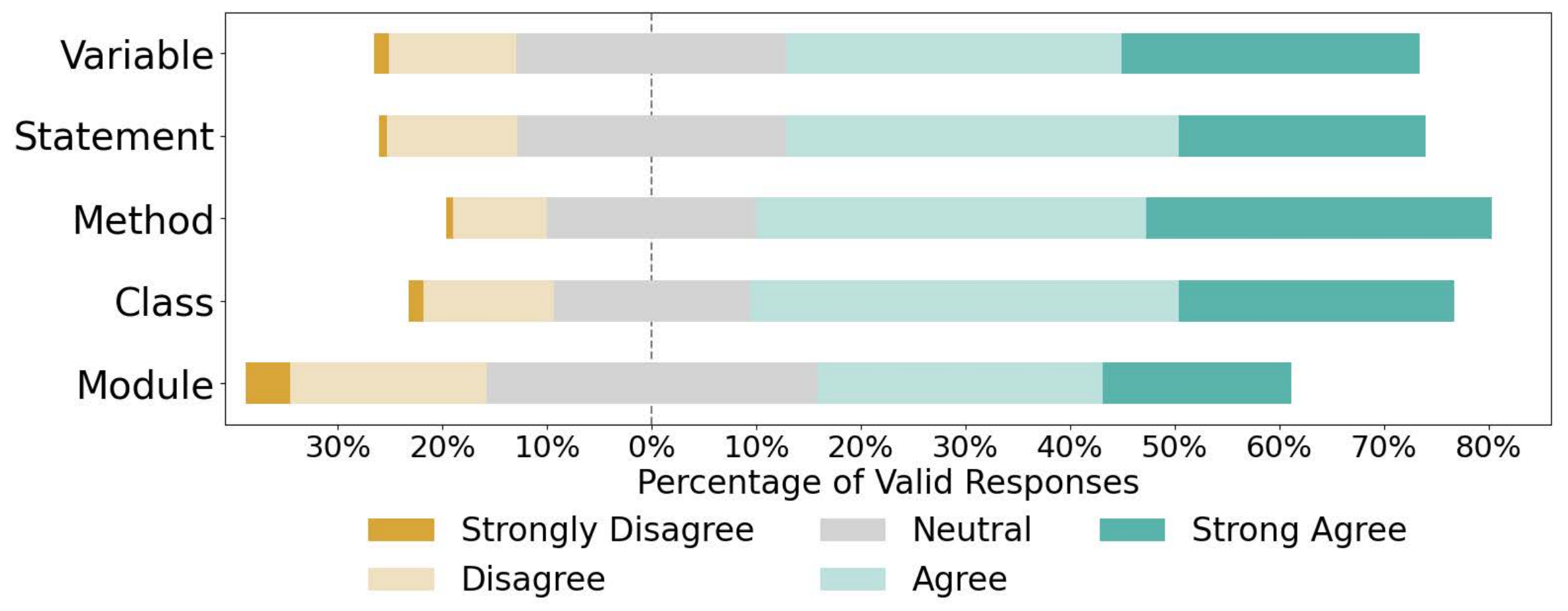}
    \caption{Respondents' Rate of Refactoring Elements' Importance}
    \label{figure-element impact}
\end{figure}

\begin{figure}
    \centering
    \includegraphics[width=.75\linewidth]{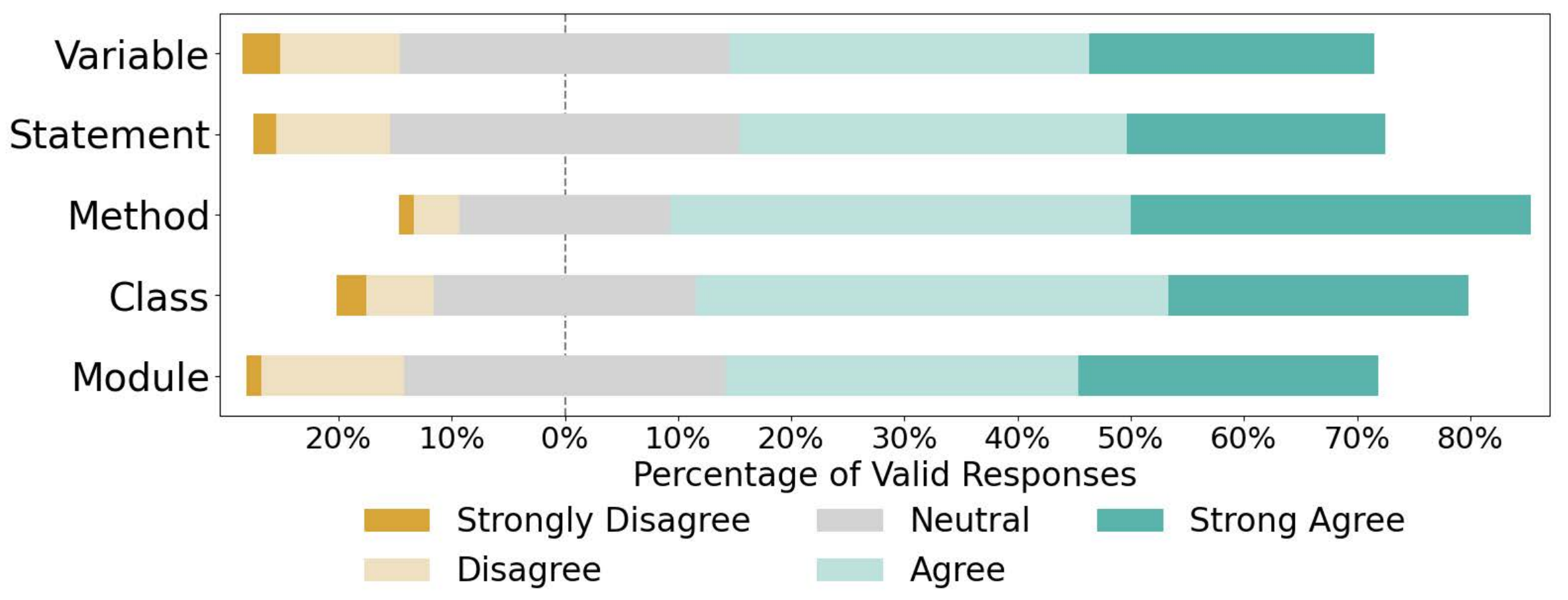}
    \caption{Respondents' Rate of Refactoring Elements' Frequency}
    \label{figure-element freq}
\end{figure}

As for refactoring elements in the context of refactoring in deep learning projects, we also invite practitioners to answer whether these refactoring operation elements are important and frequently used in deep learning projects. Figure \ref{figure-element impact}  and Figure \ref{figure-element freq} illustrate respondents' rating of refactoring elements' importance and frequency.

According to the results, \textit{Method} and \textit{Class} rank first and second, respectively, with \textit{Variable} and \textit{Statement} closely following in terms of practitioners' rating for \textit{``Do you think refactoring elements below are important to improve the quality of the code in your deep learning project?''} and \textit{``Do you think this refactoring element is frequently used in your deep learning project?''} In comparison, \textit{Module} falls significantly behind in importance score, but its rating for importance is similar to that of \textit{Variable} and \textit{Statement}.

\textit{Method} is considered crucial and frequently refactoring element in both traditional software and deep learning projects. This is consistent with our observation in RQ1. \textit{Class} comes second in rating results, which could indicate that in the area of deep learning, optimizations at the method and class level are being given priority by practitioners. They not only directly affect the logic and structure of the model but are also easier to refactor with the right level of element. \textit{Statement} and \textit{Variable} have comparable and relatively low ratings. This could imply that, for practitioners, the impact of refactoring at the statement and variable levels on code quality in deep learning projects is regarded as less significant compared to the refactoring at method and class levels. However, the frequency of variable-level refactoring usage does not align with our observations in RQ1, and a possible explanation for this could be that the minuscule of its modifications causes practitioners to neglect to carry out a Variable-level refactoring operation. \textit{Module} has a relatively low rating, this could be attributed to the fact that restructuring modules demands a comprehensive understanding of the entire project structure, which is a significant challenge for practitioners. Moreover, in deep learning projects, modules often emphasize model components, data processing flow, or training pipelines, requiring less frequent refactoring. We also find that more experienced practitioners are less likely to agree with the importance of Module-level code refactoring in deep learning projects.

\begin{center}
    \resizebox{\linewidth}{!}{
\begin{tabular}{l!{\vrule width 1pt}p{0.98\columnwidth}}
    \makecell{{\LARGE \faLightbulbO}}  &
  \textbf{Finding 5}:
\textit{\textit{Method} is seen as the most important and frequently refactored, followed by \textit{Class}. In contrast, \textit{Module} is rated lower in both importance and frequency, possibly due to its complexity and less frequent need in deep learning projects. Experienced practitioners also value \textit{Module} refactoring less.}
\end{tabular}}
\end{center}

\subsection{RQ3: Practitioners' Perspectives on Refactoring Tools}
In RQ3, we survey practitioners about their use of refactoring tools consisting of code smell detection tools and automatic refactoring tools. We further invite them to provide their perceptions of refactoring tools to uncover the shortcomings of existing tools.

\subsubsection{Code Smell Detection Tools}
Of the 159 questionnaires responding to this section, 85 respondents (53.5\%) indicate that they have used or are familiar with code smell detection tools. 73 people gave valid opinions about the tool.

There are 28 practitioners who find code smell detection tools useful but also contain many drawbacks, the biggest one is \textit{``Too many false positives''}, which is the common view of 17 practitioners. 
There are a number of concerns that have deterred practitioners from using code smell detection tools. One is \textit{``... can't figure out how to set them up...''}, and \textit{``sometimes you need to suppress the feedback and if the ability to filter out this feedback is too granular it leads to a lot of filler in your code, if it's too broad you may miss out on useful feedback.''} Some practitioners find \textit{``it difficult to distinguish more semantically
''}, and \textit{``... are often too pedantic and lack contextual information about the project structure''}.

There are also some participants who are concerned that the current tools \textit{``...are built for only traditional software eng''}, a participant finds the code smell detection tools \textit{``...usually aren't designed with ML projects in mind, and sometimes raise inappropriate warnings. For example, an ML algorithm may have many hyper-parameters, but a code smell detection tool will complain that a function shouldn't have so many arguments.''} There is also a practitioner who states \textit{``...Python is behind other languages I've used in terms of tooling for smell/refactoring''}.

In addition, many practitioners offer some insights into their go-to code smell detection tools, including SonarQube, Ruff, Flake8, IDE plugins, ChatGPT, and Copilot. SonarQube, the automated code review platform, has received a moderately positive response. 
Of the code analysis tools, practitioners seem to favour Ruff, with some practitioners arguing that \textit{``...ruff has been helpful, pylint is too hard to configure and anything more complex is too annoying...''}. One practitioner also states \textit{``Ruff is the fastest tool I used. It'll be very good if all the tools are as fast as ruff''}. Tools that combine with Large Language Models, like Copilot and ChatGPT, have recently gained popularity, that \textit{``These have been my go-to choices for code-related tasks''}.

\begin{center}
    \resizebox{\linewidth}{!}{
\begin{tabular}{l!{\vrule width 1pt}p{0.98\columnwidth}}
    \makecell{{\LARGE \faLightbulbO}}  & \textbf{Finding 6}:
\textit{  Developers acknowledge the usefulness of traditional code smell detection tools but have some concerns: (1) they always appear as false positives, (2) do not apply to deep learning projects, and (3) they are not user-friendly. Combining code smell detection tools with Large Language Models is considered the future trend.}
\end{tabular}}

\end{center}

\subsubsection{Automatic Refactoring Tools}
Of the 159 questionnaires responding to this section, 71 respondents (44.7\%) indicate that they have used or are familiar with code smell detection tools. 62 people gave valid opinions about the tool. 

There are many practitioners (23) who find automatic refactoring tools useful which is \textit{``...absolutely necessary to keep the code clean and professional''} but only for sample cases and also weak in accuracy. The automatic refactoring tool \textit{``...makes things faster but it usually doesn't handle heavy complexity too well...''}.
Similar to the code smell detection tools, practitioners find that current automatic refactoring tools are not designed for deep learning. A practitioner says that \textit{``I have found them useful for web server development mainly, but not so much for deep learning development''}. There is also a practitioner who states these tools
\textit{``...might be more challenging for loosely typed languages''}. 
The practitioners also thought these tools are hard to use, a practitioner states \textit{``The python Rope library ... docs to be somewhat lacking in ``how to'' guidance, ...documentation that serves the purposes of being detailed and then (and perhaps seperate material) is able to concisely demonstrate how to achieve refactoring operations - is needed
''}.

As a practitioner states that code smell detection tools \textit{``...Can introduce bugs or syntax issues''}, the most important concern that has deterred practitioners from working with automatic refactoring tools is their tendency to introduce bugs. Because \textit{``Sometimes wrong refactoring, takes longer to fix''}. 

The most common tools considered by practitioners are IDE plugins and tools combined with the Large Language Model, such as Copilot and ChatGPT, which are good at code generation. In contrast to traditional tools that receive a rating of \textit{``not good, not bad''}, ChatGPT and Copilot have high expectations. A practitioner states \textit{``... I have had moderately good success refactoring code with a LLM and I expect to do it more often''}. There is also a voice that the traditional automatic refactoring tools \textit{``Great but limited - integrations with Copilot would be great''}. However, a practitioner states \textit{``To really trust an automatic refactoring tool, you need to have a lot of trust in your test suite''}, and there is also a practitioner who states \textit{``...tried chatgpt and copilot. Their suggestion is the right direction to go. However, one needs to test all the edge cases thoroughly.''}

\begin{center}
    \resizebox{\linewidth}{!}{
\begin{tabular}{l!{\vrule width 1pt}p{0.98\columnwidth}}
    \makecell{{\LARGE \faLightbulbO}}  &\textbf{Finding 7}:
\textit{Developers recognize the usefulness of automatic refactoring tools. But they also have many concerns about traditional tools: (1) they only work for easy cases (2) and weak in accuracy, (3) they lack context or project structure, and (4) are hard to use. Besides, tools combined with Large Language Models have been well received, but the code they refactored is worthy of further testing.}
\end{tabular}}
\end{center}

\subsubsection{Practitioners' Advice}

There are 48 practitioners who give valid advice for refactoring tools enhancement.

The most common view is that these tools should take more information into account, including contextual and constructive information. A practitioner who offers a proposal that \textit{``...let's imagine a tool, that takes into account project decisions, rules, conventions. Moreover, tool that scans git history and `understands' such git changes that were specifically about refactorings. and let's imagine that such tool `understands' commands that are relevant to specific project in natural language format''}.
Another part of the practitioners advise that the current tools for deep learning should be enhanced with customization features. A practitioner states \textit{``More configurable so that they can also be applied to libraries and frameworks''}. A practitioner also suggested these tools should \textit{``learn the user's coding style and adapt to it instead of forcing a predefined standard''}.

The practitioners  also make their own suggestions regarding the current difficulties in using the tool.
A practitioner says the tools need \textit{``Documentation and tutorials. Additionally, having a consistent definition of operations''}. While another practitioner wants these tools \textit{``Reduce configuration variability, extend the documentation''}. There is also a part of practitioners who hope the automatic tools do not `force' them and easy to undo. A practitioner states \textit{``Sometimes I feel they `force' me to refactor my code although some parts do not need a refactoring job''}, and another practitioner says \textit{``...it should always be easy to undo/have an easily navigatable history of recent changes...''}. Furthermore, it has been suggested by numerous practitioners that refactoring must be accompanied by a testing component, as a practitioner states \textit{``The best way to refactor is by maintaining the invariant that test coverage is complete and the tests pass. test harness...''}, which is a crucial part of the refactoring process.

\begin{center}
    \resizebox{\linewidth}{!}{
\begin{tabular}{l!{\vrule width 1pt}p{0.98\columnwidth}}
    \makecell{{\LARGE \faLightbulbO}}  &\textbf{Finding 8}:
\textit{Deep learning practitioners offer their advice on how to improve current automatic refactoring tools, including (1) taking context and project structure into consideration, (2) building complete documents and configuration for ease of use, (3) adding customized features to be used in deep learning projects, (4) offering tracker of history change to rollback and (5) combining a complete test component.}
\end{tabular}}
\end{center}

\section{Discussion}
\label{sec:discussion}

\subsection{Implications}
Our results highlight a number of points to be further discussed and several implications for the research community:

\subsubsection{Strengthening research on code refactoring in deep learning projects}
 Our research demonstrates that the unique context of deep learning—such as model architectures, matrix manipulations, and hyperparameter tuning—creates distinct refactoring needs. In our dataset, \textit{Clean Up Refactoring}, \textit{Rename}, and \textit{API Refactoring} are frequent and impactful in DL projects, indicating that developers struggle most with maintaining readability and API stability during rapid prototyping.

Tang et al.~\cite{tang2021empirical} offer an initial exploration into refactoring in machine learning (ML) projects, identifying 14 “ML-specific” refactoring types from commit message patterns (e.g., ``Make algorithms more visible'' and ``Monitor feature extraction progress''). However, these ML-specific refactorings are often inferred from commit messages rather than code changes themselves, and some instances do not align strictly with standard definitions of refactoring. The underlying refactoring techniques, such as \textit{Rename}, \textit{Extract}, and \textit{Clean Up}, remain consistent with general software engineering practices but are applied in ways that address ML-specific artifacts, such as models, matrices, and feature engineering processes. By focusing on widely recognized refactoring categories, our study offers a robust foundation for understanding these shared refactoring practices in a deep learning context. Future work could further explore how these adaptations impact the effectiveness and frequency of refactoring in machine learning as the field continues to evolve.

 \subsubsection{Improving the refactoring tools for DL development}
Our findings indicate that tool developers should prioritize support for the most impactful and frequent refactoring types in DL projects, namely \textit{Clean Up}, \textit{Rename}, and \textit{API Refactoring}. These operations are essential for managing experimental code and evolving APIs during rapid prototyping.
In addition, code smell detection requires special attention in the DL context. Existing tools often enforce rigid thresholds (e.g., maximum number of parameters), which misclassify common DL practices such as large hyperparameter sets as code smells. To address this, tools should adopt project-adaptive, context-aware detection strategies rather than relying on static rules.
Finally, accuracy remains a critical concern for developers. Our survey shows that current tools frequently fail to capture the broader code context and project structure, limiting their applicability to large-scale DL systems. Enhancing detection algorithms with semantic analysis and providing explainable suggestions will improve both accuracy and developer trust, and better adoption in real-world DL projects.

 \subsubsection{Combining refactoring with testing}
 Testing is a critical aspect of refactoring, which ensures that code refactoring does not break existing functionality and maintains the stability and reliability of the system. However, to the best of our knowledge, none of the prevalent refactoring tools are currently supported by a testing system that meets the developer's requirements. Future tools should integrate lightweight, automated regression testing or leverage existing DL validation pipelines to increase developer confidence during refactoring.

\subsubsection{Refactoring in the Era of Large Language Model}
According to the respondents of our survey, current refactoring tools are \textit{``... not smart ...being inflexible''} and also lack the use of code context to support a customizable refactoring tool. By incorporating Large Language Model and code analysis techniques, automatic refactoring suggestions can be generated. This provides context-specific and personalized proposals based on the syntax rules, code context, and project structure. Tool builders should explore LLM-assisted workflows that combine static analysis with contextual reasoning to provide personalized refactoring suggestions.

\subsubsection*{Actionable Recommendations }
Based on our findings, we highlight the following implications for the community: \ding{182} Tool developers should enhance refactoring tools to support DL-specific development contexts better and explore integration with LLM-based assistants. \ding{183} Researchers are encouraged to investigate DL-specific refactoring features, particularly under dynamic typing and loose architectural conventions. 
\ding{184} DL practitioners should strengthen their understanding of refactoring principles to ensure maintainability and reusability during rapid iteration and experimentation phases, common in DL workflows.

\subsection{Threats to Validity}
\label{subsec:Threats}
Our research threats come from the two stages of our empirical study. 
                                                                                                                                                          
\subsubsection{Internal Threats}

One internal threat is the potential for missed refactoring instances due to our use of keyword-based filtering on commits. This filtering approach, while necessary to manage the data scale, might exclude relevant refactoring events, leading to an incomplete analysis. Although following prior research methods (e.g., Alomar et al.~\cite{alomar2019can}), the possibility remains that this strategy limits our refactoring detection coverage. To reduce this threat, we conduct a survey of deep learning practitioners to offer another view of real refactoring practices. Another internal threat involves the online survey stage, where participants' understanding of refactoring concepts might vary. Despite providing definitions upfront, we cannot ensure full comprehension, potentially introducing inconsistencies in participant responses. While such variability is common in studies involving practitioner insights, it may still impact the reliability of our findings.

\subsubsection{External Threats}

Our study focuses on five popular deep learning frameworks with extensive maintenance and active community involvement, which might limit the generalizability of our findings to other deep learning projects. Less mature or smaller projects may exhibit different refactoring patterns or practices. Furthermore, as our survey participants are primarily open-source contributors, there is an external validity threat regarding applicability to developers working in commercial settings. Extending this work to include commercial deep learning developers is our ongoing work, which would provide a broader perspective and enhance generalizability.



In summary, while we have taken measures to mitigate these threats where possible, the findings should be interpreted with an understanding of these limitations. Future work will aim to address these limitations by incorporating commercial project data and refining detection techniques.

\section{Related work}
\label{sec:relatedwork}

Refactoring is recognized as a fundamental practice for keeping software sustainable and healthy~\cite{silva2016we, tsantalis2020refactoringminer}. Extensive empirical research has recently been conducted to extend our knowledge of this practice. The two major lines of research related to our work are (1) studies based on refactoring practices and (2) studies based on surveys and interviews.

\subsection{Studies based on refactoring practices}

There has been much work that analyzed code changes or development documentation in repositories to gain insights related to refactoring.
Murphy-Hill et al.~\cite{Murphy-Hill_Black_Dig_Parnin_2008} investigated how developers refactor by analysing past commits. They found that programmers perform numerous refactorings within a brief time frame, and 90\% of refactorings are carried out manually.  Negara et al.~\cite{negara2013comparative} presented the first study to consider both manual and automated refactoring. It used a large set of refactorings found through an algorithm that detects them in code. Their central findings reveal that over half of the refactorings are performed manually, and 30\% of the applied refactorings do not reach the version control system. AlOmar et al.~\cite{alomar2021behind} analyzed 800 open-source projects and identified their contributors.  They found there is no correlation between experience and motivation behind refactoring, top contributed developers are found to perform a wider variety of refactoring operations. Chaves et al.~\cite{chavez2017does} analyzed the version history of 23 open-source projects with 29,303 refactoring operations. They found that developers apply more than 94\% of the refactoring operations to code elements with at least one critical internal quality attribute. Vassallo et al.~\cite{vassallo2019large} analyzed the change history of 200 open source systems' commits, and found that developers mainly schedule code refactoring after the system's structure is stable. 
Studies on ML refactoring practices are highly related to this work~\cite{tang2021empirical}.
Tang et al.~\cite{tang2021empirical} conducted a study aimed at understanding the refactoring operations specific to machine learning projects by manually mining 327 instances from 26 Java machine learning projects. They focused on unique aspects of ML code technical debt, including the transformation of primitives and parameters, algorithm visibility, and matrix operations.

There are also some studies investigating the performance of refactoring on issues consist of reuse, security, and maintenance~\cite{alomar2022refactoring,sellitto2022toward, codabuxrubbing, anthony2022refactor,lin2016interactive,pantiuchina2020developers}. Nevertheless, most of the studies above have focused on Non-deep learning projects, leaving a gap in the insights of refactoring on deep learning projects.

\subsection{Studies based on surveys and interviews}

Wang et al.~\cite{wang2009motivate} conducted a study involving 10 expert software developers. They identified both intrinsic and external factors that drive refactoring activity. The research also highlighted tool availability as a prominent factor that enables developers to translate their motivations into actions. Vakilian et al.~\cite{vakilian2012use} collected and analyzed interaction data from Java programming and found that programmers prefer lightweight refactorings and usually perform small changes using the refactoring tool. They also found that programmers use predictable automated refactorings even if they have rare bugs or change the program's behaviour.  Kim et al.~\cite{kim2014empirical} surveyed 328 engineers from Microsoft and found that developers place less importance on preserving behavior in refactoring definitions. Jain et al.~\cite{jain2019empirical} surveyed 221 IT professionals to understand the trends followed by developers and refactoring research opportunities. They found that refactoring tools are under-used as they have availability, usability, and trust issues. Oliveira et al.~\cite{oliveira2023untold} analyzed 1,162 refactorings from 13 software projects and conducted a survey with 40 developers about customization patterns. Developers confirmed the relevance of customization patterns and agreed that improvements in IDEs' refactoring support were needed. Oliveria et al.~\cite{oliveira2023towards} surveyed 53 Java developers on GitHub and found that the tools did not detect many refactorings as expected, and developers did not follow the tools' refactoring mechanisms.

The above work conducted interviews or surveys to discover developers' perspectives about refactoring and refactoring tools. However, they are still mostly limited to developers of Java projects. The findings related to refactoring in deep learning projects, which is currently a fast-growing area and is very different from non-deep learning projects in terms of development and maintenance, still need to be investigated.

\section{Conclusion and Future Work}
\label{sec:conclusion}

Code refactoring is an important part of deep learning project development. In this work, we manually analyze five deep learning projects' history commits to mine refactoring operations and further survey 159 deep learning practitioners from 38 countries for practitioners' views on refactoring and their expectations of refactoring tools. Practitioners in deep learning recognize the importance of refactoring in the development of deep learning projects. They also offer comments and advice on current refactoring tools.

We highlight the limitations of current research and suggest future directions for the improvement of code refactoring in deep learning projects. Moreover, we present practitioners' expectations regarding refactoring tools and provide suggestions for their enhancements. The complete dataset of our work can be found in the replication package~\cite{Replication_Package}.

To further improve code refactoring practices in the field of deep learning, researchers should collaborate with deep learning practitioners continuously. Future studies could put more effort into refactoring in deep learning, and develop automation tools for deep learning projects to improve the overall efficiency of code maintenance.
\section*{Acknowledgments}
This research is supported by the National Key R\&D Program of China (No. 2024YFB4506400) and the CCF-Huawei Populus Grove Fund.

\bibliographystyle{IEEEtran}
\bibliography{ICSME25}
\end{document}